# COMPARING NETWORK STRUCTURES OF COMMERCIAL AND NON-COMMERCIAL BIOHACKING ONLINE-COMMUNITIES

Sipra Bihani, Michael Hartman, Florian Sobiegalla, Amanda Rosenberg

IIT Stuart School of Business, University of Bamberg
10 West 35th Street, 18th Floor
Chicago USA, Bamberg Germany
sbihani@id.iit.edu, michael-wolfgang.hartmann@stud.uni-bamberg.de, florian.sobiegalla@me.com,
arosenberg@id.iit.edu

**ABSTRACT**
This paper compares two biohacking groups, Bulletproof Executive and DIYbio, whose distinct goals result in differences in social network structures, activities and entry points.

**INTRODUCTION**
The term biohacking was first used online in December 2008 and now encompasses a wide range of topics and activities (Google Trends). We define biohacking as the amateur practice of biological experimentation for a self-defined purpose using a variety of DIY devices and techniques in a non-traditional setting (Bennett 2009; Delfanti 2013; personal communication, October 29 2014).

The two groups this paper analyzes are DIYbio and Bulletproof Executive (BE.) DIYbio was founded in 2008 by Mackenzie Cowell and Jason Bobe with the mission of "establishing a vibrant, productive and safe community of DIY biologists" (DIYbio). BE is led by founder Dave Asprey, who has a personal and financial interest in discovering methods that help people "reach the state of high performance" (Bulletproof). These two groups have grown their networks online in different ways.

**COMPARISON OF BIOHACKING GROUPS**
Both BE and DIYbio use online social media to share knowledge, expand their network, and meet their respective goals. Both groups have a website, a blog, a Facebook page, and a Twitter account; DIYbio also has a newsletter and Google group while BE has multiple Twitter accounts, a YouTube channel, an Instagram account, a Google+ page, a podcast, and an online store. The five methods used to develop the comparison are secondary research for context on biohacking and trends; an interview an active biohacker who attended the 2014 Bulletproof Conference to learn the diversity of motivations of biohackers; a survey to DIYbio local chapters to understand their definition of biohacking and current activities; and social network analysis with Condor, Gephi, and Wordle to create a network of actors; and content analysis with TagCrowd of the two group's blogs to analyze activities.

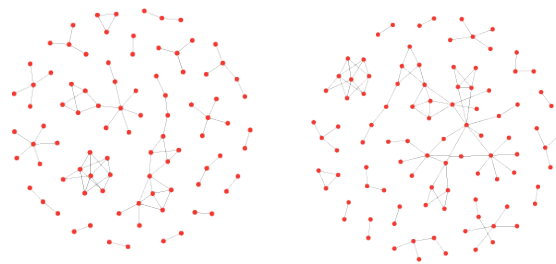

*Figure 1 (left) : Twitter fetch for "DIYbio" on Dec 6, 2014; Figure 2 (right): Twitter fetch for "LondonBioHack" on Dec 6, 2014*

DIYbio has the mission of democratizing biological research, which has implications for its network structure and entry points. As Figure 1 shows, the network has no central node. The network is composed of 55 regional COINs from 21 countries that set their own protocols (DIYbio). Figure 2 shows the tweets of "LondonBioHack", which is a local chapter. Actors tweeting about this group are more connected with one another than those tweeting about DIYbio. Since actors create new technology and ideas, LondonBioHack and DIYbio both coolfarm with projects such as OpenerPCR and an outline of biohacking ethics, respectively (survey, Nov 19, 2014, DIYbio, Gloor 2010).

In regards to entry points, DIYbio has no form of public outreach, such as advertising, for people outside of the network to learn about it. The local groups get public visibility by disseminating their findings on their chapter websites and through members' personal blogs. In accordance with Gloor, DIYbio can be classified as a self-selecting collaborative innovation network (COIN) because members aim to contribute to current scientific knowledge in an open source environment (2007, 2010). People who are or want to become a part of the virtual network are intrinsically motivated to reach a shared goal and develop their own set of rules.

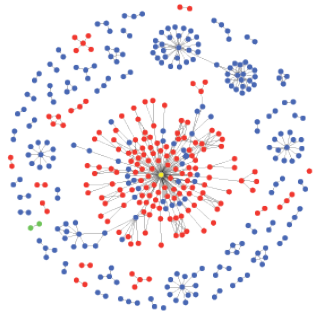

*Figure 3: Twitter Fetch for "Bulletproof Coffee" in blue, "Bulletproofexec" in red, "Dave Asprey" in green on Nov. 19, 2014. The central yellow node is (@bulletproofexec.)*

To examine the structure of BE, we looked at its two Twitter accounts: @bulletproofexec and @BPNutrition. As Figure 3 shows, Asprey, as represented by the handle @bulletproofexec, is a central node in the BE network; @BPNutrition is a bridge to other accounts. There are tweets about Bulletproof Coffee by accounts that are not directly connected to either Asprey or @BPNutrition. The commercial product has spurred attention to form a Collaborative Interest Network (CIN) with members who share a common interest and practice without necessarily identifying as biohackers (Gloor, 2010).

Asprey continues to be a coolhunter by brokering a network with different biohacking circles and experts on self-improvement topics that he brings together on his social media and at the annual Bulletproof Conference (Gloor 2007). In an interview with a participant at the conference, the biohacker said that he felt like it was the first real biohacking conference because it covered a variety of topics such as electron flow, gut ecology, habit formation, and nutrient injections (personal communication, October 29 2014.) This brokering enables BE to be influential in many biohacking topic areas and provides channels for commercialized products.

Through its different social media outlets, BE promotes the company's central product and entry point, called Bulletproof Coffee, which is marketed as more effective at improving productivity and efficiency than the average cup of coffee. People without a specific interest in biohacking can learn about Bulletproof Coffee by stumbling upon it at a local coffee shop or by reading about it in media outlets, such as The New York Times (Bulletproof). As Figure 4 shows, the easily identifiable Bulletproof Coffee had growth in Google searches that have far surpassed searches for related terms. Based on the commercial entry point and the mix of biohackers and non-biohackers, BE can be described as a COIN with a CIN in which actors share a common interest and spot new trends that will become cool (Gloor 2007).

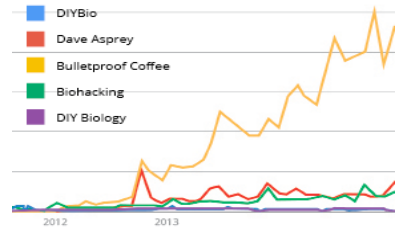

*Figure 4: Google Trend visualization of interest in search terms from January 2008 to December 2014.*

## CONCLUSION

The research finds a significant interrelation among organization goals, structure, activity, and entry points as shown in Table 1. DIYbio is an interconnected network of deeply engaged biohackers that coolfarms, whereas, by coolhunting and forming CINs, BE exposes many people to biohacking methods and products for self-improvement.

|  | BE | DIYbio |
|---|---|---|
| Network structure | a COIN with a CIN | a COIN of COINs |
| Network influence | leader | no leader |
| Activity | coolhunting | coolfarming |
| Entry points | commercial | open-source |

*Table 1: Comparison of BE and DIYbio based on influence, structure, focus, and entry points.*

Since biohacking is a young and growing practice, some questions remain about the social network structures of the two organizations under examination. As the communities mature, their structure may evolve. For instance, will the CIN members drinking Bulletproof Coffee enter the COIN and begin biohacking? Will the umbrella organization of DIYbio continue to connect networks or will they become more and more autonomous?